\documentclass[prc,showpacs,showkeys,superscriptaddress,twocolumn]{revtex4}
\usepackage{graphicx,mathrsfs,amsmath,amssymb,amsthm,amsopn,mathbbol,bm} 
\usepackage[dvips]{color}
\def\cal{\mathcal}

\let\sectiontmp\section\let\subsectiontmp\subsection \let\appendixtmp\appendix
\def\section{\setcounter{equation}{0}\sectiontmp}
\def\subsection{\subsectiontmp}
\def\theequation{\arabic{section}.\arabic{equation}}
\def\appendix{\def\theequation{\Alph{section}.\arabic{equation}}\appendixtmp}
\usepackage{feynmp}
\usepackage{graphicx}
\def\scr#1{\mbox{\scriptsize #1}}
\newcommand{\Tr}{{\mathrm{Tr}}}
\def\ii{\mathrm{i}}
\def\mt{\tilde{m}}
\def\Mt{\tilde{M}}

\begin{document}
\title{
Gapless Hartree--Fock Resummation Scheme for the $O(N)$ Model} 

\author{Yu.B.~Ivanov}\thanks{e-mail: Y.Ivanov@gsi.de}
\affiliation{Gesellschaft
 f\"ur Schwerionenforschung mbH, Planckstr. 1,
D-64291 Darmstadt, Germany}
\affiliation{Kurchatov Institute, Kurchatov
sq. 1, Moscow 123182, Russia}
\author{F. Riek}\thanks{e-mail: F.Riek@gsi.de}
\affiliation{Gesellschaft
 f\"ur Schwerionenforschung mbH, Planckstr. 1,
D-64291 Darmstadt, Germany}
\author{J. Knoll}\thanks{e-mail: J.Knoll@gsi.de}
\affiliation{Gesellschaft
 f\"ur Schwerionenforschung mbH, Planckstr. 1,
D-64291 Darmstadt, Germany}


\begin{abstract}
  A modified self-consistent Hartree--Fock approximation to the
  $\lambda\phi^4$ theory with spontaneously broken $O(N)$ symmetry is
  proposed. It preserves all the desirable features, like conservation
  laws and thermodynamic consistency, of the self-consistent
  Schwinger-Dyson scheme generated from a 2PI functional, also known
  as the $\Phi$-derivable scheme, while simultaneously respecting the
  Nambu--Goldstone theorem in the chiral-symmetry broken phase.
  Various approximate
  resummation schemes are discussed. 
\end{abstract}
\date{\today}
\maketitle

\begin{fmffile}{phi-func}
\def\fmfsdot#1{\fmfv{decor.shape=circle,decor.filled=full,decor.size=1.4thick}
{#1}}
\def\fmffdot#1{\fmfv{decor.shape=circle,decor.filled=full,decor.size=2.4thick}
{#1}}
\def\fmfcross#1#2{
\fmfv{decor.shape=cross,decor.angle=#1,decor.size=5thick}{#2}
}
\def\PhiF{
\parbox{14mm}{\centerline{
\begin{fmfgraph*}(10,10)
\fmfpen{thick}
\fmfleft{l}
\fmfright{r}
\fmftop{t}
\fmfbottom{b}
\fmfforce{(0.5w,0.5h)}{m}
\fmf{plain}{t,b}
\fmf{plain}{l,r}
\fmfsdot{m}
\fmfcross{0}{b,l,r,t}
\end{fmfgraph*}}
}}
\def\JPhiF{
\parbox{14mm}{\centerline{
\begin{fmfgraph*}(10,10)
\fmfpen{thick}
\fmfleft{l}
\fmfright{r}
\fmftop{t}
\fmfbottom{b}
\fmfforce{(0.5w,0.5h)}{m}
\fmf{plain}{t,b}
\fmf{phantom}{l,ll}
\fmf{plain}{m,r}
\fmf{plain,width=1}{ll,m}
\fmffdot{m}
\fmfcross{0}{b,r,t}
\end{fmfgraph*}}
}}
\def\PhiTtad{
\parbox{12mm}{\centerline{
\begin{fmfgraph*}(10,10)
\fmfpen{thick}
\fmfleft{lb,lt}
\fmfforce{0.5w,1.3h}{t}
\fmfright{rb,rt}
\fmfforce{(0.5w,0.5h)}{m}
\fmf{plain}{lb,m,rb}
\fmf{plain,right=1}{m,t,m}
\fmfsdot{m}
\fmfcross{45}{lb,rb}
\end{fmfgraph*}}
}}
\def\JPhiTtad{
\parbox{12mm}{\centerline{
\begin{fmfgraph*}(10,10)
\fmfpen{thick}
\fmfleft{lb,lt}
\fmfforce{0.5w,1.3h}{t}
\fmfright{rb,rt}
\fmfforce{(0.5w,0.5h)}{m}
\fmf{phantom}{lb,llb}
\fmf{plain,width=1}{llb,m}
\fmf{plain}{m,rb}
\fmf{plain,right=1}{m,t,m}
\fmffdot{m}
\fmfcross{45}{rb}
\end{fmfgraph*}}
}}
\def\SPhiTtad{
\parbox{12mm}{\centerline{
\begin{fmfgraph*}(10,10)
\fmfpen{thick}
\fmfleft{lb,l,lt}
\fmfforce{0.5w,1.3h}{t}
\fmfright{rb,r,rt}
\fmfforce{(0.5w,0.5h)}{m}
\fmf{plain,width=1}{ll,m,rr}
\fmf{plain}{lb,m,rb}
\fmf{phantom,tension=2}{l,ll}
\fmf{phantom,tension=2}{r,rr}
\fmffdot{m}
\fmfcross{45}{lb,rb}
\end{fmfgraph*}}
}}
\def\Phieight{
\parbox{12mm}{\centerline{
\begin{fmfgraph*}(10,16)
\fmfpen{thick}
\fmfleft{lt,lb}
\fmfforce{0.5w,0h}{b}
\fmfforce{0.5w,1h}{t}
\fmfright{rt,rb}
\fmfforce{(0.5w,0.5h)}{m}
\fmf{plain,right=1}{m,t,m}
\fmf{plain,right=1}{m,b,m}
\fmfsdot{m}
\end{fmfgraph*}}
}}
\def\Phisun{
\parbox{25mm}{\centerline{
\begin{fmfgraph*}(20,10)
\fmfpen{thick}
\fmfleft{l}
\fmfforce{0.2w,0.5h}{ll}
\fmfforce{0.8w,0.5h}{rr}
\fmfright{r}
\fmf{plain}{l,ll,rr,r}
\fmf{plain,right=0.75}{ll,rr,ll}
\fmfsdot{ll,rr}
\fmfcross{0}{l,r}
\end{fmfgraph*}}
}}
\def\Phifootball{
\parbox{18.5mm}{\centerline{
\begin{fmfgraph*}(16,25)
\fmfpen{thick}
\fmfleft{l}
\fmfright{r}
\fmf{plain,right=0.34}{l,r,l}
\fmf{plain,right=0.8}{l,r,l}
\fmfsdot{l,r}
\end{fmfgraph*}}
}}
\def\SPhieight{
\parbox{12mm}{\centerline{
\begin{fmfgraph*}(10,16)
\fmfpen{thick}
\fmfleft{lt,l,lb}
\fmfforce{0.5w,0h}{b}
\fmfforce{0.5w,1h}{t}
\fmfright{rt,r,rb}
\fmfforce{(0.5w,0.5h)}{m}
\fmf{plain,right=1}{m,t,m}
\fmf{plain,width=1}{ll,m,rr}
\fmf{phantom,tension=2}{l,ll}
\fmf{phantom,tension=2}{r,rr}
\fmffdot{m}
\end{fmfgraph*}}
}}
\def\JPhisun{
\parbox{25mm}{\centerline{
\begin{fmfgraph*}(20,10)
\fmfpen{thick}
\fmfleft{l}
\fmfforce{0.2w,0.5h}{ll}
\fmfforce{0.8w,0.5h}{rr}
\fmfright{r}
\fmf{plain,width=1}{l,ll}\fmf{plain}{ll,r,rr}
\fmf{plain,right=0.75}{ll,rr,ll}
\fmffdot{ll}
\fmfsdot{rr}
\fmfcross{0}{r}
\end{fmfgraph*}}
}}
\def\SPhisun{
\parbox{25mm}{\centerline{
\begin{fmfgraph*}(20,10)
\fmfpen{thick}
\fmfleft{lb,l,lt}
\fmfforce{0.2w,0.5h}{ll}
\fmfforce{0.8w,0.5h}{rr}
\fmfright{rb,r,rt}
\fmf{plain}{lt,ll}\fmf{plain}{rr,rt}
\fmf{plain,width=1}{ll,l}\fmf{plain,width=1}{r,rr}
\fmf{plain,right=0.75}{ll,rr,ll}
\fmffdot{ll,rr}
\fmfcross{-65}{lt}\fmfcross{65}{rt}
\end{fmfgraph*}}
}}
\def\SPhisunsp{
\parbox{25mm}{\centerline{
\begin{fmfgraph*}(20,10)
\fmfpen{thick}
\fmfleft{lb,l,lt}
\fmfforce{0.2w,0.5h}{ll}
\fmfforce{0.8w,0.5h}{rr}
\fmfright{rb,r,rt}
\fmf{plain}{lt,ll}\fmf{plain}{rr,rt}
\fmf{plain,width=1,label=$\pi$,label.side=left}{ll,l}
\fmf{plain,width=1,label=$\pi$,label.side=left}{r,rr}
\fmf{plain,right=0.75,label=$\pi$}{ll,rr}
\fmf{plain,right=0.75,label=$\sigma$}{rr,ll}
\fmffdot{ll,rr}
\fmflabel{$\sigma$}{lt}\fmflabel{$\sigma$}{rt}
\fmfcross{-65}{lt}\fmfcross{65}{rt}
\end{fmfgraph*}}
}}
\def\SPhisunt{
\parbox{35mm}{\centerline{
\begin{fmfgraph*}(33,10)
\fmfpen{thick}
\fmfleft{lb,l,lt}
\fmfforce{0.1w,0.5h}{ll}
\fmfforce{0.9w,0.5h}{rr}
\fmfforce{0.5w,0.5h}{mm}
\fmfright{rb,r,rt}
\fmf{plain}{lt,ll}\fmf{plain}{rr,rt}
\fmf{plain,width=1}{ll,l}\fmf{plain,width=1}{r,rr}
\fmf{plain,right=0.75}{ll,mm,ll}
\fmf{plain,right=0.75}{rr,mm,rr}
\fmfsdot{mm}
\fmffdot{ll,rr}
\fmfcross{-65}{lt}\fmfcross{65}{rt}
\end{fmfgraph*}}
}}
\def\SPhifootball{
\parbox{25mm}{\centerline{
\begin{fmfgraph*}(20,25)
\fmfpen{thick}
\fmfleft{l}
\fmfforce{0.2w,0.5h}{ll}
\fmfforce{0.8w,0.5h}{rr}
\fmfright{r}
\fmf{plain}{ll,rr}
\fmf{plain,width=1}{l,ll}\fmf{plain,width=1}{r,rr}
\fmf{plain,right=0.75}{ll,rr,ll}
\fmffdot{ll,rr}
\end{fmfgraph*}}
}}
\section{Introduction}

The $O(N)$ model with the spontaneously broken symmetry is a
traditional touchstone for new theoretical approaches with
applications to a variety of physical phenomena, such as the
chiral phase transition in nuclear matter \cite{Gell-Mann,Pisarski},
the formation of Bose--Einstein condensates in atomic gases
\cite{Wachter,Andersen}, or inflationary cosmology scenario
\cite{Kofman,Lee}. However, even for such a simple model, conceptual
problems arise for the dynamic and even thermodynamic treatment of
the system. To be precise, the present discussion concerns
non-perturbative approaches based on partial resummation schemes of
various kinds: the $\Phi$-derivable approximation of Baym \cite{Baym}
or CJT formalism \cite{Cornwall} as it is referred in the field
theory, $1/N$ expansions constructed on top of the CJT formalism
\cite{Berges01,Berges02},
gapless approximations\footnote{This notion reflects the fact that the
  massless Goldstone boson has a gapless spectrum.} \cite{HM65},
approximations of mixed type like the variational
two-particle-point-irreducible (2PPI) concepts \cite{Verschelde,Baacke} or
the Baym--Grinstein approximation \cite{Baym-Grin}.

{Central objective of this paper is to construct self-consistent
  dynamical equations of motion for the mean-field $\phi(x)$ and
  one-particle propagator $G(x,y)$ with improved features with respect
  to symmetry properties. The space-time changes of both quantities
  can be formulated as a coupled set of mean-field and
  Schwinger-Dyson (SD) equations in their differential form
  \begin{eqnarray}\label{gen-dyn-eq}
    {G}_0^{-1}\odot
\phi(x) &=& J[\phi,G],
\\
    {G}_0^{-1}\odot
G(x,y) &=& I +\Sigma[\phi,G]\odot G.
    \label{gen-dyn-eq1}
 \end{eqnarray}
 Here ${G}_0^{-1}$ is the inverse free propagator, the unity $I$ and the
 $\odot$-operation are to be understood in the corresponding
 functional sense. The self-consistence of the scheme requires both
 source terms, the current $J$ and the self-energy $\Sigma$, to be
 {\em entirely} functionals of $\phi$ and $G$. Baym \cite{Baym,KB}, see
 also \cite{IKV98a}, has shown that such a scheme is conserving at the
 expectation value level and at the same time thermodynamically
 consistent {\em if and only if} both driving terms are derived from
 a single functional $\Phi[\phi,G]$ through functional variations
  \begin{eqnarray}
\label{Phi-der-J}
    J &=&\frac{\delta}{\delta \phi} \Phi[\phi,G],
\\
\label{Phi-der-S}
    -i \Sigma_{12}&=& \frac{\delta}{\delta G_{21}}\Phi[\phi,G].
  \end{eqnarray}
  Approaches of such $\Phi$-derivable type permit truncations of the
  $\Phi$-functional at any level, this way providing a variety of
  approximations without violating the above features. Equipped with
  the required irreducibility features, i.e. that the closed diagrams
  of $\Phi$ are two-particle irreducible (2PI), any such truncation
  scheme, through the coupled set of equations
  (\ref{gen-dyn-eq}) and (\ref{gen-dyn-eq1}), provides
  a partial resummation of diagrams to infinite order, which is void
  of double counting and with built-in detailed balance. In its Wigner
  transformed version it leads to the well known Kadanoff-Baym
  equations \cite{KB} which through gradient approximation permits the
  derivation of generalized transport
  equations \cite{IKV99,IKV01b,IKV02}. 
}

{ Despite all the above features, however, $\Phi$-derivable
  approximations in the sense of Eqs
  (\ref{gen-dyn-eq})--(\ref{Phi-der-S}) violate Ward--Takahashi
  identities beyond the 
  one-point level.  In particular, this leads to a violation of the
  Nambu--Goldstone (NG) theorem \cite{Baym-Grin,HM65,HK3} in case of
  spontaneously broken symmetry (which is referred to as the
  Hugenholtz--Pines theorem in the physics of Bose--Einstein
  condensation).  The standard remedy is to construct a 1PI effective action
  $\Gamma\left[G_\Phi [\phi],\phi\right]$ based on the chosen 
$\Phi$-derivable approximation, i.e. $G_\Phi [\phi]$ being a solution
  of Eq. (\ref{gen-dyn-eq}) with arbitrary field $\phi$. The functional
  variation of this effective action with respect to $\phi$ results in
  equations for higher order vertices, 
} 
e.g. the Bethe--Salpeter (BS) equation. This BS-equation
(the RPA equation in the case of the HF approximation) has to be
solved supplementary to the SD equation with {\em frozen} propagators
of the SD scheme (\ref{gen-dyn-eq}), cf.  Ref. \cite{HK3}. The thus
constructed vertex functions, in particular the RPA self energy, do
indeed respect the corresponding Ward--Takahashi identities and hence
the NG theorem.  However, these quantities do not take part in the
self-consistent scheme in the sense 
{ that they do not enter back
  into the source terms in Eq. (\ref{gen-dyn-eq}). In turn, the {\em
    internal} propagators entering in (\ref{gen-dyn-eq}) are still not
  gapless and still suffer from violating the NG theorem.
In view of non-equilibrium applications, where all
  dynamical quantities are evolved in time, this is an unacceptable
  deficiency of the remedy scheme via higher-order vertex
  functions. Moreover, the corresponding BS equation numerically requires
  huge computing resources once one includes higher order
  diagrams in the $\Phi$ functional.  Generally this remedy scheme,
  constructed ``on-top'' of a self-consistent $\Phi$-derivable
  SD approximation, is of a ``gapless'' type in terms of
  classification of Ref.  \cite{HM65}.  This gapless approximation
  respects the NG theorem, yet violates conservation laws and
  thermodynamic consistency \cite{HM65}.  }

In this respect, the particular statement in Ref. \cite{Nemoto} and
some subsequent publications that ``the NG theorem is always satisfied
at any finite $N$ in the Hartree--Fock approximation'' of the
$\Phi$-functional (or CJT) formalism, is certainly misleading. This
statement is based on the identification of the pion and sigma masses
with the curvatures of the effective potential at its minimum.
{These curvature masses are actually precisely those given by
  the RPA scheme.  They are again constructed ``on
  top'' of the self-consistent scheme as they nowhere re-enter in the
  driving terms of the coupled meand-field-SD scheme
  (\ref{gen-dyn-eq})--(\ref{gen-dyn-eq1}). 
What however matters in self-consistent
  dynamical treatments, both for transport and thermodynamics, are the
  pole masses of the self-consistent Green functions.
}  This fact was
realized long ago in the theory of Bose--Einstein condensation
\cite{HM65,Wachter}. As a consequence, kinetic descriptions of
condensed systems were based on the gapless approximations
\cite{HM65,Wachter} rather than the $\Phi$-functional, accepting the
lack of conservation laws as a minor defect compared to the else
poorly reproduced Goldstone boson.

The above statement can also formulated in formal terms. Strictly
speaking, the NG theorem \cite{Goldstone} claims that the inverse
propagator of the pion vanishes as $p^2$ as $p^2\to 0$. If the Green
function is calculated as the second variation of the effective 
action over fields (like in the ``gapless scheme''), then the zero
curvature of the effective 
potential at its minimum in the ``pion direction'' is equivalent to
the the NG theorem. However, this is precisely the problem that in the
$\Phi$-functional (or CJT) formalism the 
Green function is associated with stationary point of the effective 
action with respect to the variation over  the Green function $G$ itself
rather than the fields. In this case, the fact that the curvature of
the effective potential at its minimum is zero is irrelevant to the NG
theorem. This is why the recovery of the ``gapless scheme'' was
the main idea behind the restoration of the NG theorem, reported in
Ref. \cite{HK3}. In fact, similar arguments on the validity of the
NG theorem have been already put forward in Ref. \cite{Cooper05}.

From this point of view, the
two-particle-point-irreducible\footnote{graphs that fall apart, if two
  line meeting at the same point, are cut; in the Hartree--Fock
  approximation it is identical to the standard 2PI condition.} (2PPI)
approach \cite{Baacke} still violates the NG theorem.  In leading
(zero) order in the $1/N$ expansion the $\lambda \phi^4$ model is
indeed free of the above-discussed problems.  However, with $N=2$ for
Bose--Einstein condensation and $N=4$ for the chiral phase transition
in nuclear matter this zero order scheme is far from being adequate.
Already terms of next-to-leading order in $1/N$ in the self-energies
violate the NG theorem in the above sense, i.e. due to the finite pole
mass of the pion, contrary to the statement of Ref.  \cite{Berges02}.
This is clearly seen in the thermodynamic limit of this formalism
\cite{Cooper05}.

First attempts to find a compromise between $\Phi$-derivable and
gapless schemes were undertaken by Baym and Grinstein
\cite{Baym-Grin}.  In fact, their modified Hartree--Fock approximation is a
simplified gapless scheme, which still respects the NG theorem.
Therefore it was accepted as a main tool for studies of 
disoriented chiral
condensates \cite{Anselm,Blaizot,Randrup}. Baym and Grinstein
\cite{Baym-Grin} also found severe problems with the renormalization
of partial resummation schemes. Great progress in the proper
renormalization of such schemes was recently achieved in Refs
\cite{HK3,HK1,HK2,Reinosa1,Reinosa2}.  However, we will not touch the
question of renormalization in this article, since it leads to extra
complications which deserve a special discussion.

In this paper we discuss the standard example of a spontaneously
broken $O(N)$ model in $\Phi$-derivable Hartree--Fock
approximation\footnote{This approximation is also frequently referred
  as the Hartree approximation, e.g. see \cite{Lenaghan}. In fact,
  exchange terms, i.e. the Fock terms, are included into this
  approximation, therefore the term ``Hartree--Fock'' is more precise.
  For contact interaction the distinction between direct and exchange
  terms is not so obvious. It becomes visible, if one replaces the
  contact interaction by a heavy-boson exchange interaction. In the
  $1/N$ counting scheme the Fock terms prove to be of subleading order
  \cite{Aouissat}.}. The considerations are confined to the
thermodynamic equilibrium, since the discussed problems reveal
themselves already at that level. Our goal is to construct a modified
Hartree--Fock approximation, which is still $\Phi$-derivable, and
{thus automatically provides conservation laws and
  thermodynamically consistence at the level of the corresponding
  mean-field--Schwinger-Dyson scheme (\ref{gen-dyn-eq}), and at the same time
  {\em respects the NG theorem}}. The present treatment is based on a
naive renormalization, where all divergent terms are simply omitted.

\section{Conventional Hartree--Fock  Approximation}

We consider the $O(N)$-model Lagrangian 
\begin{eqnarray}
\begin{split}
{\cal L} =& \frac{1}{2} (\partial_\mu \phi_a)^2 - \frac{1}{2} m^2 \phi^2
- \frac{\lambda}{4N} (\phi^2)^2 \\&+ H\cdot \phi, 
\end{split}
\end{eqnarray}
where $\phi = (\phi_1,\phi_2,...,\phi_N)$ is an $N$-component scalar
field, $\phi^2 = \phi_a \phi_a$ with summations over $a$ implied. For
$H=0$ this Lagrangian is invariant under $O(N)$ rotations of the
fields. If $H=0$ and $m^2 < 0$, the symmetry of the ground state is
spontaneously broken 
down to $O(N-1)$, with $N-1$ Goldstone bosons (pions). The external
field $H\cdot \phi = H_a \phi_a$ is a term which explicitly breaks the
$O(N)$ symmetry.  It is introduced to give the physical
value of 140 MeV to the pion mass.

The conventional Hartree--Fock  approximation to the $O(N)$ model is
defined by the $\Phi$ functional\footnote{All the considerations below
  are performed in terms of the thermodynamic $\Phi$ functional which
  differs from the real-time $\Phi$ in the factor of $\ii\beta$, where
  $\beta=1/T$ is the inverse temperature. In the case of a spatially
  homogeneous thermodynamic system, an additional factor appears: the
  volume $V$ of the system. Thus, our $\Phi=(-\ii
  T/V)\Phi_{\scr{real-time}}$. This $\Phi$ is similar to the $V_2$
  part of the effective potential in the CJT formalism \cite{Cornwall}
  with the exception that, contrary to $V_2$, 
by definition $\Phi$ includes all 2PI
  interaction terms, i.e. also those of zero and first loop order
  which result from interactions with the classical field (first two
  graphs in (\ref{Phi-Hartree})).}
\begin{eqnarray}
\label{Phi-Hartree}
\Phi_{\scr{HF}}=\PhiF+\PhiTtad+\Phieight, 
\end{eqnarray}
where the crosses denote the classical field $\phi$.  Within the
$\Phi$-derivable scheme the r.h.s. of the equations of motion for the
classical field ($J$) and the Green function (self-energy $\Sigma$) follow
from the functional variation of $\Phi_{\scr{HF}}$ with respect
to the classical field $\phi$ and Green function $G$, respectively
\begin{eqnarray}
\label{HF}\label{Phi-J}\label{Phi-S}
\begin{split}
\Box\phi+m^2\phi=&J=\displaystyle 
\frac{\delta\Phi_{\scr{HF}}}{\delta\phi}\\\\=&\JPhiF+\JPhiTtad,\\[5mm]
G^{-1}_{\scr{HF}}-D^{-1}=&
\Sigma_{\scr{HF}}=\displaystyle 
2 \frac{\delta\Phi_{\scr{HF}}}{\delta G}
\\\\=&\SPhiTtad+\SPhieight, 
\end{split}
\end{eqnarray}
where $D$ is the free propagator. In the spontaneously broken-symmetry
phase, i.e. 
for $H=0$ and $m^2<0$, the solutions of equation set (\ref{HF})
violate the NG theorem, as it has been demonstrated in numerous papers
(see e.g.  Refs \cite{Baym-Grin,Lenaghan,Aouissat}). A detailed
analysis of these equations in a notation similar to ours has been given
in Ref.  \cite{Lenaghan}.

It is possible to remedy the defect of the Hartree--Fock  approximation by
correcting the self-energy by standard random phase approximation (RPA) 
technique as follows \cite{HK3,Aouissat} 
\begin{eqnarray}
\label{gapless}
\Sigma_{\scr{gapless}}
=\SPhiTtad +\SPhieight +\Delta\Sigma^{\scr{RPA}},
\end{eqnarray}
where 
\begin{eqnarray}
\begin{split}
\label{deltaS}
\Delta\Sigma^{\scr{RPA}} =& \SPhisun \\\\&+ \SPhisunt + ... 
\end{split}
\end{eqnarray}
This $\Delta\Sigma^{\scr{RPA}}$ is constructed in terms of Green
functions $G_{\scr{HF}}$ obtained from solving the
self-consistent Hartree--Fock  equations (\ref{HF}) and in fact
represents the RPA series \cite{Aouissat}.  Actually, this correction
converts the scheme into that of gapless rather than of
$\Phi$-derivable type. This results from the loss of the
self-consistency, since the 
RPA self-energies do not enter the SD-equation (\ref{Phi-S}).  The
most important ingredient of this scheme is the pion self-energy
\begin{eqnarray}
\begin{split}
\label{deltaSp1}
\Delta\Sigma^{\scr{RPA}}_\pi &= \SPhisunsp + ...\\\\
&= \left(\frac{\lambda}{N}\right)^2 \sigma^2 
\frac{(\ii G_\pi \ii G_\sigma)}%
{1 - \frac{2\lambda}{N}(\ii G_\pi \ii G_\sigma)}
\end{split}
\end{eqnarray}
which provides the NG theorem. Here
\begin{eqnarray}
\begin{split}
\label{(GG)}
(\ii G_\pi \ii G_\sigma) &\\&\equiv 
\int_\beta d^4 q \ii G_\pi(k+q/2) \ii G_\sigma(k-q/2), 
\end{split}
\end{eqnarray}
where the Matsubara integration 
\begin{eqnarray}
\begin{split}
\label{int-T}
\int_\beta d^4 q f(q) &\\&\equiv T \sum_{n=-\infty}^\infty 
\int \frac{d^3 q}{(2\pi)^3} f(2\pi\ii nT,\vec{q}), 
\end{split}
\end{eqnarray}
is implied with $T$ being a finite temperature. At $H=0$, 
by means of the Hartree--Fock equations of motion (\ref{HF}), 
$\Delta\Sigma^{\scr{RPA}}_\pi(k=0)$
is reduced to \cite{Aouissat} 
\begin{eqnarray}
\label{deltaSpk0}
\Delta\Sigma^{\scr{RPA}}_\pi(k=0) = -
\frac{2\lambda}{N} (Q_\pi - Q_\sigma), 
\end{eqnarray}
where 
\begin{eqnarray}
\label{Pnr}
Q_a 
\equiv 
\int_\beta d^4 k G_a(k) 
\end{eqnarray}
is the tadpole diagram of the Green function
\begin{eqnarray}
\label{G-H}
G_a(k) = \frac{1}{k^2+M_a^2},   
\end{eqnarray}
which is the solution of the Hartree--Fock equations of motion (\ref{HF})
with $M_a$ being the effective mass  of the particle. 
Upon substitution to
(\ref{gapless}), this $\Delta\Sigma^{\scr{RPA}}_\pi$ provides
$G^{-1}_{\pi\scr{(gapless)}}(k=0)=0$ or 
$M^2_{\pi\scr{(gapless)}}=M^2_\pi+\Delta\Sigma^{\scr{RPA}}_\pi(k=0)=0$ 
in the broken-symmetry phase. This is precisely the requirement of the NG
theorem. 

However, the recovery of the NG theorem is achieved here on the
expense of loss of self-consistency (i.e. $\Phi$-derivable nature) of
the approximation. It means that the dynamic treatment based on scheme
(\ref{gapless}) will not possess proper conservation laws and
thermodynamic consistency is lost. The latter means in particular,
that the numbers of particles calculated as a derivative of the
thermodynamic potential $\Omega$
\begin{eqnarray}
\begin{split}
\label{Omega}
\frac{1}{T}\Omega\{\phi,G\}=&I_0(\phi)+
\frac{1}{2}\Tr\left(\ln G^{-1}\right)\\
&-\frac{1}{2}\Tr\left(\Sigma G\right)+\Phi\{\phi,G\}
\end{split}
\end{eqnarray}
with $I_0(\phi)$ being the free classical action of $\phi$ field, over
chemical potential\footnote{of course, provided we introduce a
  chemical potential}, on the one side, and as the Green function $G$
integrated over momentum, on the other side, will be different. 
Moreover, the spectral function of the sigma meson calculated
within this approach has the wrong threshold behavior \cite{HK3},
which however does not affect the validity of the NG theorem. 
It is not gapless because of the still nonzero pole mass of the pion
Green function (\ref{G-H}). Moreover, the chiral phase transition still proves
to be of the first rather than of the second order, as required by the
universality class of the $O(N)$ model.

\section{Gapless Hartree--Fock (gHF) Approximation}

For the exact theory, i.e. when all diagrams of the $\Phi$ functional
are taken into account, both the gapless and $\Phi$-derivable schemes
become identical. It is desirable to have them identical also at a
certain approximation level such as the Hartree--Fock approximation. If
not completely identical, then at least identical in the
Goldstone-boson sector, which is of prime importance.  The latter
means that corrections to the pion self-energy
(\ref{deltaSp1})  primarily need to be incorporated in a $\Phi$-derivable way.
In fact, we need far less to satisfy the NG theorem, namely only that
the pion self-energy vanishes at vanishing four-momentum $p$.

For this purpose one can  introduce {\em a phenomenological
symmetry-restoring correction to the $\Phi$ functional} of
Eq. (\ref{Phi-Hartree}) 
\begin{eqnarray}
\label{dPhi}
\Delta\Phi = -
\frac{(N-1)\lambda}{2N} (Q_\pi - Q_\sigma)^2
\end{eqnarray}
which precisely produces the required correction term, namely the RPA
term $\Delta\Sigma^{\scr{RPA}}_\pi(k=0)$ to the pion
self-energy, see Eq. (\ref{deltaSpk0}). This $\Delta\Phi$ can be
presented in a manifestly $O(N)$-symmetric form 
\begin{eqnarray}
\label{dPhi-ab}
\Delta\Phi = -
\frac{\lambda}{2N} \left(N Q_{ab}Q_{ba} - Q_{aa}Q_{bb}\right),
\end{eqnarray}
where summations over $a$ and $b$ are implied and 
\begin{eqnarray}
\label{Pab}
Q_{ab} = \int_\beta d^4 k G_{ab}(k)
\end{eqnarray}
is the tadpole diagram of the Green function $G_{ab}$. Thus, the
$\Phi$ functional taken as a functional of the classical field $\phi$
and the Green functions $G_{ab}$ for the gapless Hartree--Fock approximation
is defined as
\begin{eqnarray}
\label{Phi-eff}
\Phi_{\scr{gHF}}[\phi,G_{ab}] =
\Phi_{\scr{HF}}[\phi,G_{ab}] + \Delta\Phi [G_{ab}]  
\end{eqnarray}
In the broken-symmetry phase with non-vanishing classical field
$\phi=\{\phi_a\}$ 
and in an arbitrary, not necessarily diagonal, representation the
Green functions, $G_{ab}$ takes the form
\begin{eqnarray}
\label{Gab_H}
G_{ab}=\frac{\phi_a\phi_b}{\phi^2} G_\sigma +
\left(\delta_{ab} - \frac{\phi_a\phi_b}{\phi^2}\right) G_\pi 
\end{eqnarray}
in terms of the pion and sigma propagators, cf. e.g.\cite{Baacke}. 

The advantage of the $\Phi_{\scr{gHF}}$ is that the 
resulting $\Phi$-derivable approximation obeys the NG theorem, at the
same time keeping all the pleasant features of the $\Phi$-derivable
approach. Moreover, $\Phi_{\scr{gHF}}$ does not change
the Hartree--Fock equation for the classical field $\phi$ (\ref{HF}), as
$\Delta\Phi$ does not depend on $\phi$, cf. Eq. (\ref{Phi-J}). This is a
desirable feature, since the classical field equation coincides in both
the $\Phi$-derivable and the gapless schemes, and therefore it is reasonable
to keep this unchanged also in the gapless Hartree--Fock approximation.
Another advantage of $\Delta\Phi$ is that it does not change the
results in the $O(N)$-symmetry restored phase, where $M^2_\sigma =
M^2_\pi$ and hence $Q_\pi = Q_\sigma$ and $\Delta\Sigma_a =0$. Indeed,
this phase is described quite reasonably within conventional Hartree--Fock
approximation and hence requires no modifications. For the further
discussion of this modified $\Phi$-derivable approximation we use the 
notion in terms of CJT effective potential, see e.g.
\cite{Cornwall,Nemoto,Lenaghan}, in order to comply with numerous
previous considerations in the literature.

The manifestly symmetric form of the CJT effective potential in
the conventional Hartree--Fock approximation reads
\begin{widetext}
\begin{eqnarray}
\label{V-H}
V_{\scr{HF}}(\phi,G) &=& \frac{1}{2} m^2 \phi^2 + 
\frac{\lambda}{4N} (\phi^2)^2 - H\cdot \phi +
\frac{1}{2} \int_\beta d^4 k \ln\det G^{-1}(k)
\cr
&+&
\frac{1}{2} \int_\beta d^4 k \left\{
\left[\left(k^2+m^2\right)\delta_{ab} + \frac{\lambda}{N} 
\left(\phi^2 \delta_{ab} + 2 \phi_a\phi_b\right)
\right] G_{ba}(k)-1\right\}
%
\cr
&+&
\frac{\lambda}{4N} 
\left(Q_{aa} Q_{bb} + 2 Q_{ab}Q_{ba} \right),
\end{eqnarray}
\end{widetext}
e.g., cf. \cite{Nemoto}. 
The phenomenological symmetry-restoring correction, corresponding to
$\Delta\Phi$ of Eq. (\ref{dPhi}), is 
\begin{eqnarray}
\label{dV-H}
\Delta V = -
\frac{\lambda}{2N} \left(N Q_{ab}Q_{ba} - Q_{aa}Q_{bb}\right), 
\end{eqnarray}
where 
\begin{eqnarray}
\label{Gab}
G_{ab}^{-1}(k) = k^2\delta_{ab} + M^2_{ab}
\end{eqnarray}
is the Green function, and $Q_{ab}$ 
is the corresponding tadpole, cf. (\ref{Pab}). Here and below,
summation over repeated 
indices $a,b,c,...$ is implied, if it is not pointed out otherwise. 
All the quantities are symmetric with respect to permutations of these
indices.

The equations for $G_{cd}$, i.e. for the corresponding tadpoles $Q_{ab}$,
and the fields $\phi_c$ result from variations of
$V_{\scr{gHF}}=V_{\scr{HF}}+\Delta V $ over $G_{cd}$ and $\phi_c$,
respectively,
\begin{widetext}
\begin{eqnarray}
\label{modH-SD}
M^2_{cd} &=& m^2\delta_{cd} + \frac{\lambda}{N} \left[\phi^2 \delta_{cd} 
+ 2 \phi_c\phi_d +3 Q_{aa} \delta_{cd} + 2(1-N) Q_{cd}\right], 
\\
\label{modH-MF}
H_c &=& m^2\phi_c + \frac{\lambda}{N} \left[\phi^2\phi_c
+ Q_{aa}\phi_c + 2 Q_{cd} \phi_d\right].
\end{eqnarray}
These equations are in an arbitrary non-diagonal representation. Introducing projectors on $\pi$ and $\sigma$ states
\begin{eqnarray}
\label{pi-proj}
\Pi^\pi_{cd}&=&\frac{1}{N-1}\left(\delta_{cd} - \phi_c\phi_d/\phi^2\right),
\\
\label{sig-proj}
\Pi^\sigma_{cd}&=&\phi_c\phi_d/\phi^2, 
\end{eqnarray}
one arrives at 
\begin{eqnarray}
\label{modH-SD-pi}
M^2_\pi &=:& \Pi^\pi_{dc}M^2_{cd} = m^2 + \frac{\lambda}{N} 
\left[\phi^2 + Q_{aa} 
+ 2 \left(\phi_a\phi_b/\phi^2\right)Q_{ba}\right], 
\\
\label{modH-SD-sig}
M^2_\sigma &=:& \Pi^\sigma_{dc}M^2_{cd} = m^2 + \frac{\lambda}{N} 
\left[3\phi^2 +3 Q_{aa} 
+ 2(1-N) \left(\phi_a\phi_b/\phi^2\right)Q_{ba}\right] 
\end{eqnarray}
from  Eq. (\ref{modH-SD}). In order to project the mean-field equation (\ref{modH-MF}) on the
$\sigma$-direction, we just multiply it by $\phi_c$
\begin{eqnarray}
\label{modH-MF-sig}
 \phi^2\left\{m^2 + \frac{\lambda}{N} \left[\phi^2
+ Q_{aa} + 2 \left(\phi_a\phi_b/\phi^2\right)Q_{ba}\right]\right\} = H_c\phi_c.
\end{eqnarray}

\newpage\end{widetext}
In the diagonal representation ($\phi_\sigma\ne 0$, $H_\sigma = H$ and
$H_\pi=\phi_\pi=0$) these equations take the following form 
\begin{eqnarray}
\label{Phi-SD-s}
M^2_\sigma &=& m^2 + \frac{\lambda}{N} \left[3 \phi^2
+ (5-2N) Q_\sigma + 3(N-1) Q_\pi\right] 
\cr &=&
M^2_\pi + \frac{\lambda}{N} \left[2 \phi^2
+ 2(N-1) (Q_\pi - Q_\sigma)\right]
\\
\label{Phi-SD-p}
M^2_\pi &=& m^2 + \frac{\lambda}{N} \left[\phi^2
+ 3 Q_\sigma + (N-1) Q_\pi\right]  
\\
\label{MF-eq}
H &=&\phi \left[ m^2 + \frac{\lambda}{N} \left(\phi^2
+ 3 Q_\sigma + (N-1) Q_\pi\right)\right].  
\end{eqnarray}
Here we used
$Q_\sigma=Q_{\sigma\sigma}$ and $Q_\pi=Q_{\pi\pi}$ in terms of
definition (\ref{Pab}). The tadpoles consist of two parts: 
$Q_a = Q_a^T + Q_a^{\scr{(div)}}$, one of them 
\begin{eqnarray}
\label{P-div}
Q_a^{\scr{(div)}} = \int \frac{d^3 k}{(2\pi)^3} \frac{1}{(k^2+M_a^2)^{1/2}} 
\frac{1}{2}.
\end{eqnarray}
is divergent. Here we adopt a naive renormalization, which consists in
omitting this divergent contribution. Therefore, the tadpoles are
defined as follows 
\begin{eqnarray}
\label{P}
Q_a\equiv Q_a^T &=& \int \frac{d^3 k}{(2\pi)^3} 
\frac{1}{(k^2+M_a^2)^{1/2}} \\
&&\times\frac{1}{\exp[(k^2+M_a^2)^{1/2}/T]-1} .\nonumber
\end{eqnarray}
From these equations it is evident that the NG theorem is
fulfilled. Indeed, in the broken-symmetry phase ($H=0$) the square-bracketed
term of the field equation (\ref{MF-eq}) equals zero, which precisely
determines the pion mass, cf. Eq. (\ref{Phi-SD-p}).

\subsection{Vacuum ($T = 0$)}
\label{Vacuum}

At $T=0$, the quantities under investigation are ``experimentally''
known\footnote{These values are relevant to the case $N=4$.}: 
$M_\pi(T=0)=m_\pi=$ 139 MeV,  $M_\sigma (T=0)=m_\sigma=$ 600 
MeV, and the pion decay constant $\phi_0=f_\pi=$ 93 MeV. These known
quantities should satisfy Eqs. (\ref{Phi-SD-s})--(\ref{MF-eq})
at $T = 0$, i.e. with  $Q_\sigma=Q_\pi=0$. 
In order to make this set of equations consistent, we should put
\begin{eqnarray}
\label{H-pi}
H=m^2_\pi f_\pi, 
\end{eqnarray}
i.e. precisely the same as at the  tree level. Then equations
(\ref{Phi-SD-p}) and (\ref{MF-eq}) become identical. Resolving
Eqs. (\ref{Phi-SD-s})--(\ref{MF-eq})
with respect to $m^2$ and $\lambda$ in terms of
``experimental'' quantities $m_\pi$,  $m_\sigma$ and 
$f_\pi$, we arrive at 
\begin{eqnarray}
\label{lambda}
\hspace*{-5mm}
\lambda &=&
\frac{N\left(m^2_\sigma - m^2_\pi \right)}{2f_\pi^2},  
\\
\label{mass}
\hspace*{-5mm}
-m^2 &=& -m^2_\pi 
+
\frac{\lambda}{N} f_\pi^2.
\end{eqnarray}

\subsection{Symmetry Restoration Point $T_1$ (at $H=0$)}

If $H=0$, above some temperature the $O(N)$ symmetry is restored. 
First, we apply the strong condition
\begin{eqnarray}
\label{c_ponit}
M^2_\sigma = M^2_\pi = \phi^2 = 0. 
\end{eqnarray}
These equations determine  the symmetry restoration temperature $T_1$. 
In this case all three equations
(\ref{Phi-SD-s})--(\ref{MF-eq}) reduce to the same single one with the
solution   
\begin{eqnarray}
\label{Tc}
T_1^2 &=& \frac{12}{(N+2)} f_\pi^2. 
\end{eqnarray}
Let us consider the region in the
vicinity of $T_1$, assuming $|T-T_1|\ll T_1$, $\phi\ll T_1$ and $M_a\ll T_1$. 
Here we use brief notation
$$ \mt^2 = - m^2/T^2 >0,
\quad s=\phi/T ,\quad \Mt_a = M_a/T, $$
$$ R = -\mt^2 + \frac{\lambda}{N} (N+2)\frac{1}{12}
=  
\frac{\lambda}{N} \frac{1}{12} (N+2)
\frac{T^2-T_1^2}{T^2}. $$
Keeping only linear in $\Mt_a$ terms in
Eqs (\ref{Phi-SD-s})--(\ref{MF-eq}) accordingly to (\ref{P20}), we
arrive at the following set of equations 

\begin{figure*}[ht]
\begin{center}
\vspace*{5mm} 
\includegraphics[height=6cm,width=5cm]{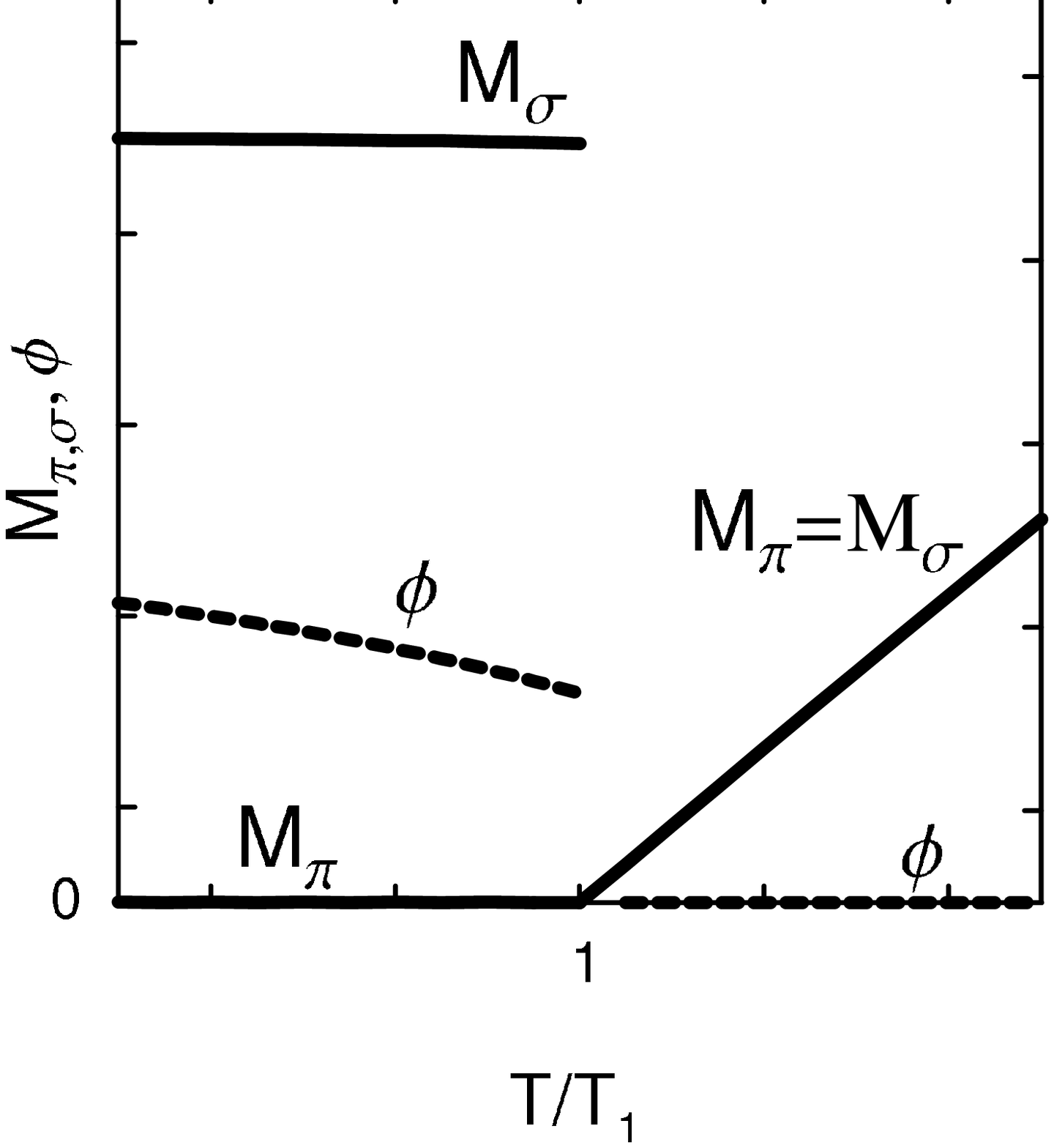}
\hspace*{9mm}
\includegraphics[height=5.7cm,width=4cm]{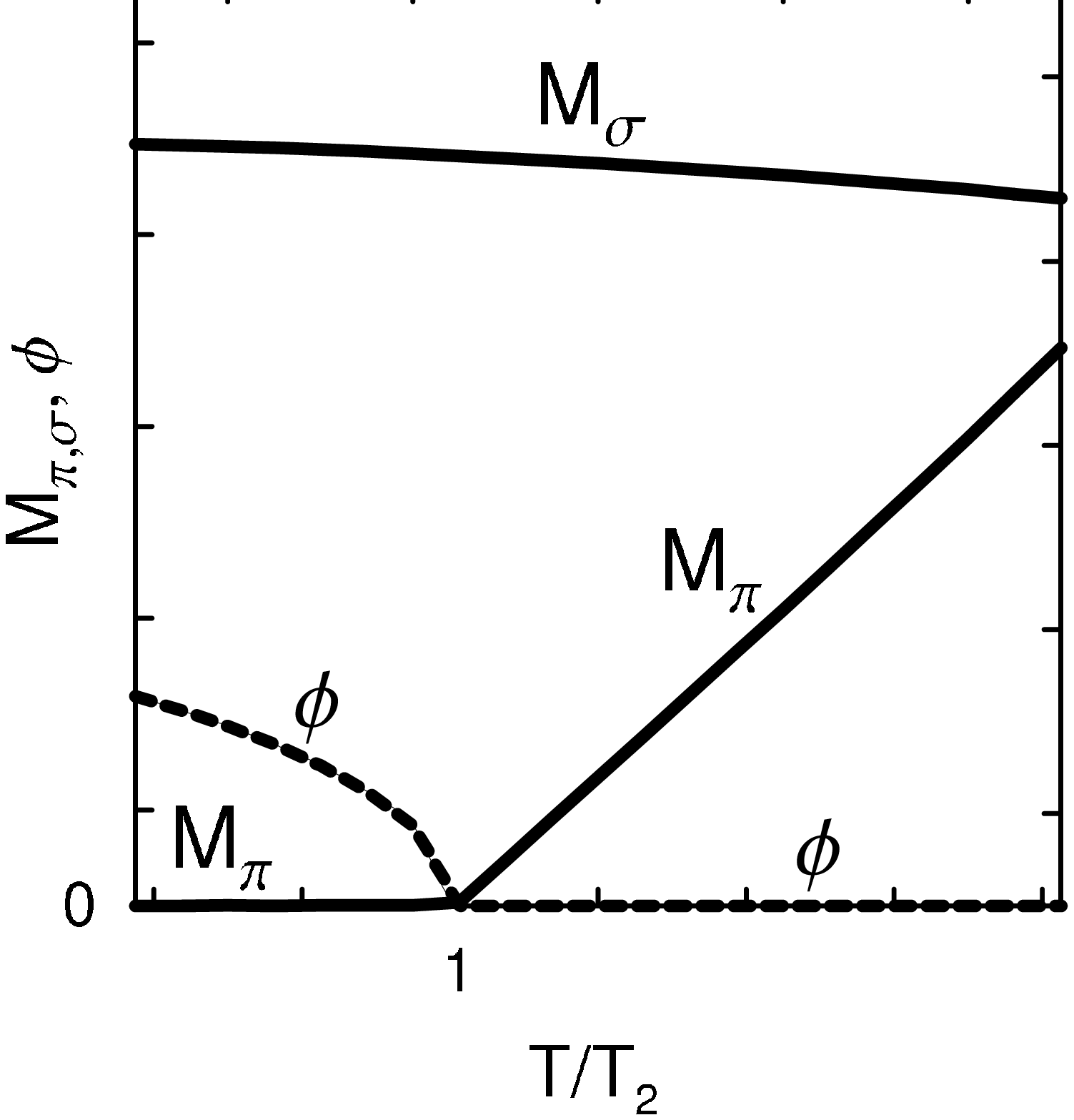} 
\vspace*{3mm} 
\caption{Schematic behavior of the solution in the vicinity of the
  $T_1$ (left panel) and $T_2$ (right panel) points.} 
\end{center} 
\label{fig-t1}
\end{figure*}

\begin{widetext}
\begin{eqnarray}
\label{Phi-SD-sc1}
0 &=& 
\frac{\lambda}{N} \left[2 s^2
+
\frac{2(N-1)}{4\pi} \left(
\Mt_\sigma - \Mt_\pi\right)
\right],
\\
\label{Phi-SD-pc1}
0 &=& R + \frac{\lambda}{N} \left(s^2
-
\frac{1}{4\pi} \left[ 3 \Mt_\sigma + (N-1)  \Mt_\pi 
\right]
\right),
\\
\label{MF-eqc1}
0 &=& s \left[ 
R + \frac{\lambda}{N} \left(s^2
-
\frac{1}{4\pi} \left[ 3 \Mt_\sigma + (N-1)  \Mt_\pi 
\right]
\right)\right]. 
\end{eqnarray}
\end{widetext}
The solution to this set with $s=0$ is trivial 
\begin{eqnarray}
\label{sol-sym}
s&=&0,\cr \Mt_\sigma &=&  \Mt_\pi = \frac{N}{\lambda}\frac{4\pi}{(N+2)}R\\
&=& \frac{\pi}{3} \frac{T^2-T_1^2}{T^2}\nonumber
\end{eqnarray}
and exists only at $T>T_1$. This is the phase of restored symmetry. 
If $s\ne 0$, then $\Mt_\pi=0$. Hence we arrive at the solution 
\begin{eqnarray}
\label{Tc-s}
s^2 &=& - \frac{N}{\lambda}\frac{2(N-1)}{2N+1} R\cr
&=&- \frac{2(N-1)}{2N+1}\frac{1}{12} (N+2) \frac{T^2-T_1^2}{T^2}, 
\\
\label{Tc-X}
\Mt_\sigma &=& \frac{N}{\lambda}\frac{8\pi}{2N+1} R\cr
&=&\frac{8\pi}{2N+1}\frac{1}{12} (N+2) \frac{T^2-T_1^2}{T^2}, 
\end{eqnarray}
implying that there are no solutions with $s^2>0$ and $\Mt_\sigma>0$
simultaneously. 
In fact, this solution exists only at $T>T_1$, since $\Mt_\sigma$
should be positive, as required by definition of $Q^T_\sigma$.
However, at $T>T_1$, the solution is unphysical, since $s^2$ becomes
negative: $s^2<0$. Of course, there exists a finite solution with
$\phi>0$ and $M_\sigma>0$ at $T<T_1$, as it is numerically
demonstrated in the next section.  This behavior of the solution in
the vicinity of $T=T_1$ is schematically demonstrated in Fig. 1.  It
implies that the phase transition at $T_1$, if it ever occurs, is of
the first order, since the physical quantities reveal jumps at this
point.%

\begin{figure*}[ht]
\vspace*{5mm} 
\begin{center}
\includegraphics[width=5.5cm,angle=-90]{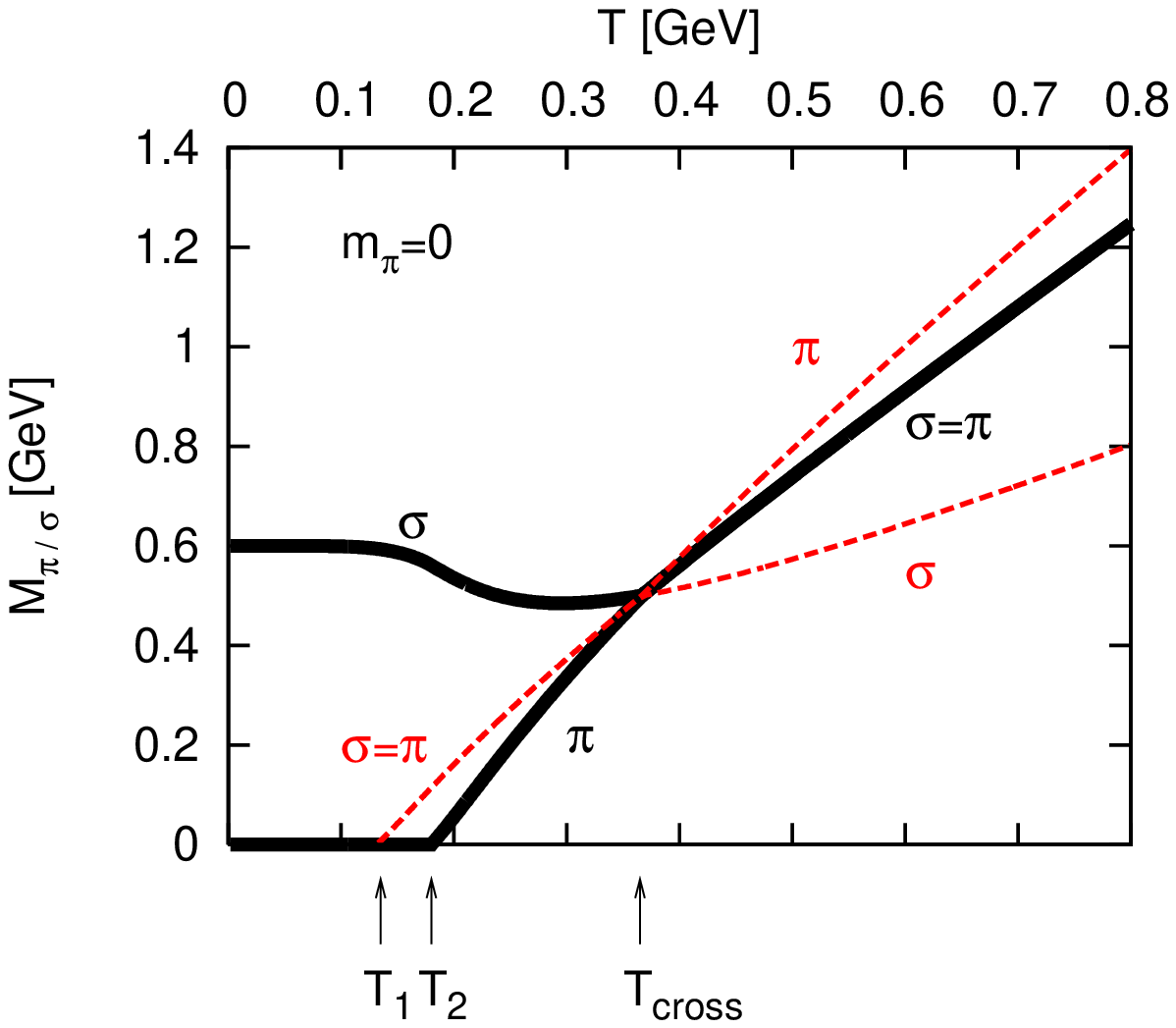}
\includegraphics[width=5.5cm,angle=-90]{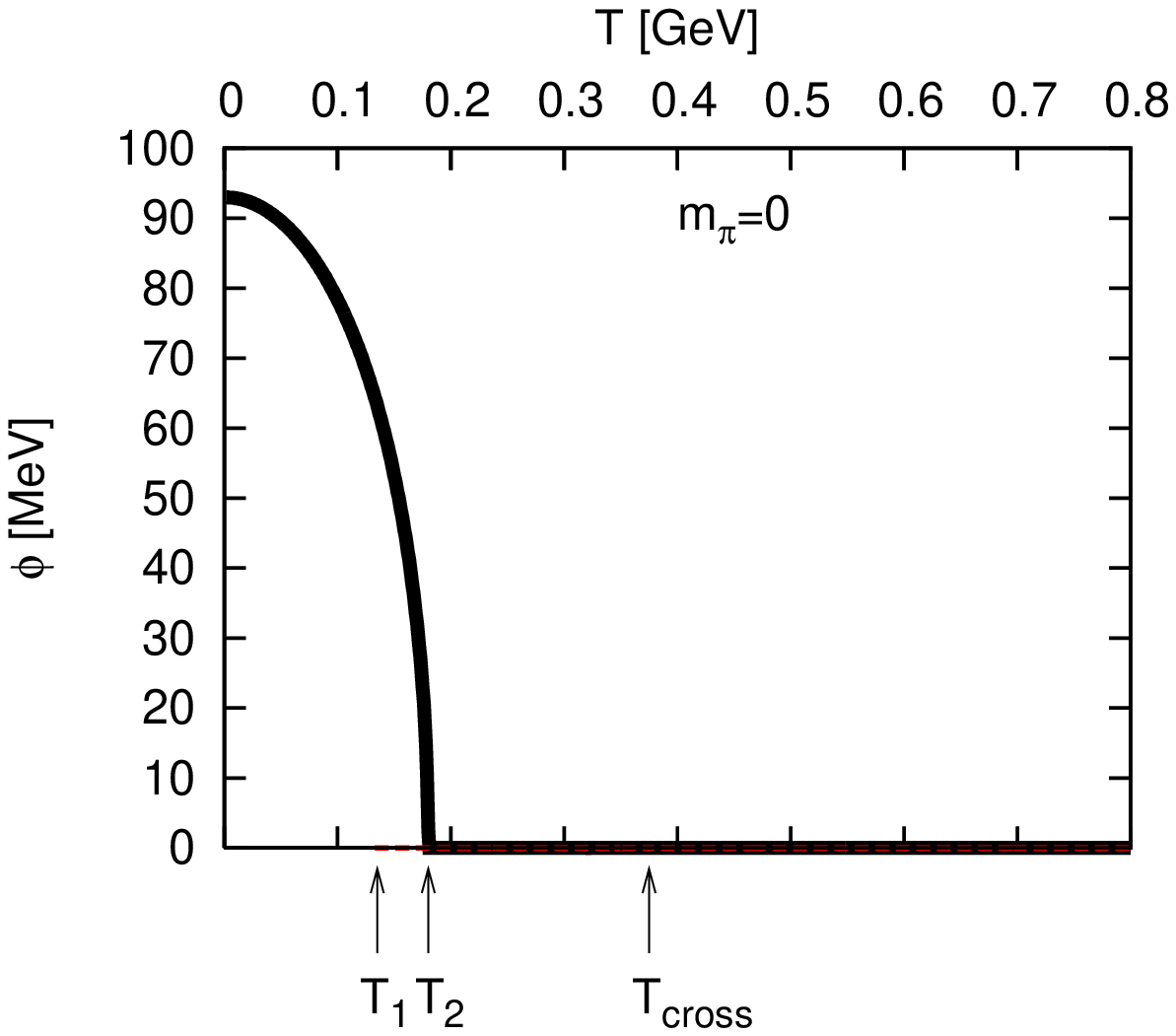}
\caption{Meson masses (left panel) and $\phi$ (right panel) as 
  functions of temperature for the $m_\pi=0$ case. Stable and 
  metastable branches are presented by solid and dashed lines,
  respectively.}
\end{center}
\label{fig-m-mp0}
\end{figure*}
\subsection{Partial Symmetry Restoration Point $T_2$ (at $H=0$)}

The transition point $T_2$ is determined by the weak condition
\begin{eqnarray}
\label{c2_ponit}
M^2_\pi = \phi^2 = 0, \quad M^2_\sigma\ne 0.
\end{eqnarray}
The symmetry is not completely
restored at this point, since the meson masses still remain different.  
A calculation very similar to that above 
gives 
\begin{eqnarray}
\label{T2}
T_2^2 = \frac{12 f_\pi^2}{(N+2)} 
\left(1+\frac{3}{(N-1)}\frac{M_\sigma^2(T_2)}{m_\sigma^2}\right)
\end{eqnarray}
Strictly speaking, this is not a solution for $T_2$, since the
r.h.s. of (\ref{T2}) still depends on $T_2$ through
$M_\sigma^2(T_2)$. However, it numerically occurs that  
$M_\sigma(T_2)\approx m_\sigma$, which almost removes this
dependence, and hence  
\begin{eqnarray}
\label{T2appr}
T_2^2 \simeq \frac{12}{(N-1)} f_\pi^2
\end{eqnarray}
The behavior of the solution in the vicinity of the $T_2$ is
schematically demonstrated in Fig. 1. 
Such a behavior implies that the phase transition at $T_2$, if it ever
occurs, is of the second order, since quantities reveal no jumps and
only their derivatives are discontinues at this point.

\section{Results for $N=$ 4}

The actual structure of the solution to Eqs
(\ref{Phi-SD-s})--(\ref{MF-eq}) turns out to be even more involved
than qualitatively discussed above.  This is demonstrated for the case
of exact $O(4)$ symmetry, i.e. $H=0$, in Fig. 2.

First of all, there are two different  solutions to
the set of Eqs. (\ref{Phi-SD-s})--(\ref{MF-eq}). 
To decide which branch is more stable, we should
compare the corresponding values of the CJT effective potential,
cf. Eqs (\ref{V-H}) and (\ref{dV-H}), which in explicit form reads
\begin{widetext}
\begin{eqnarray}
\label{Veff}
V_{\scr{gHF}} &=& 
\frac{1}{2} m^2 \phi^2 + \frac{\lambda}{4N} \phi^4 - H \phi +
L_\sigma + (N-1) L_\pi 
\cr
&+&
\frac{1}{2} 
\left(m^2 + \frac{3\lambda}{N} \phi^2 - M^2_\sigma\right) Q_\sigma 
+ \frac{1}{2} (N-1)
\left(m^2 + \frac{\lambda}{N} \phi^2 - M^2_\pi\right)Q_\pi 
\cr
&+&
\frac{\lambda}{4N} 
\left[(5-2N) Q_\sigma^2 +  (N-1)^2 Q_\pi^2 +  6(N-1)Q_\sigma Q_\pi\right], 
\end{eqnarray}
where 
\begin{eqnarray}
\label{La}
L_a = \frac{T}{(2\pi)^3}\int d^3 k
\ln\left(1-\exp\frac{(k^2+M_a^2)^{1/2}}{T}\right)  
\end{eqnarray}
\end{widetext}
and $Q_a$ are given by Eq. (\ref{P}) with $M_a$ being solutions of
equations of motion.  
Note also that this effective potential is nothing else but the negative
pressure  
\begin{eqnarray}
\label{pressure}
P = -V_{\scr{gHF}}.   
\end{eqnarray}
A stable branch should correspond to a minimum of the effective
potential with respect to the field variation, i.e. the second
derivative $d^2V_{\scr{gHF}}(\phi,G(\phi))/d\phi^2$ should be
positive. As it is seen from Eqs (\ref{d2Veff})--(\ref{R}), all
branches with positive masses $M_\sigma$ and $M_\pi$ correspond
to minima. This is precisely the case for all branches
discussed here. This minimum is only very shallow, when $M_\pi=0$.
Therefore, we can only choose between stable and metastable
branches. The most stable branch should have minimal absolute value of
$V_{\scr{gHF}}$, i.e. maximal pressure.


\begin{figure*}[ht]
\begin{center}
\includegraphics[width=5.5cm,angle=-90]{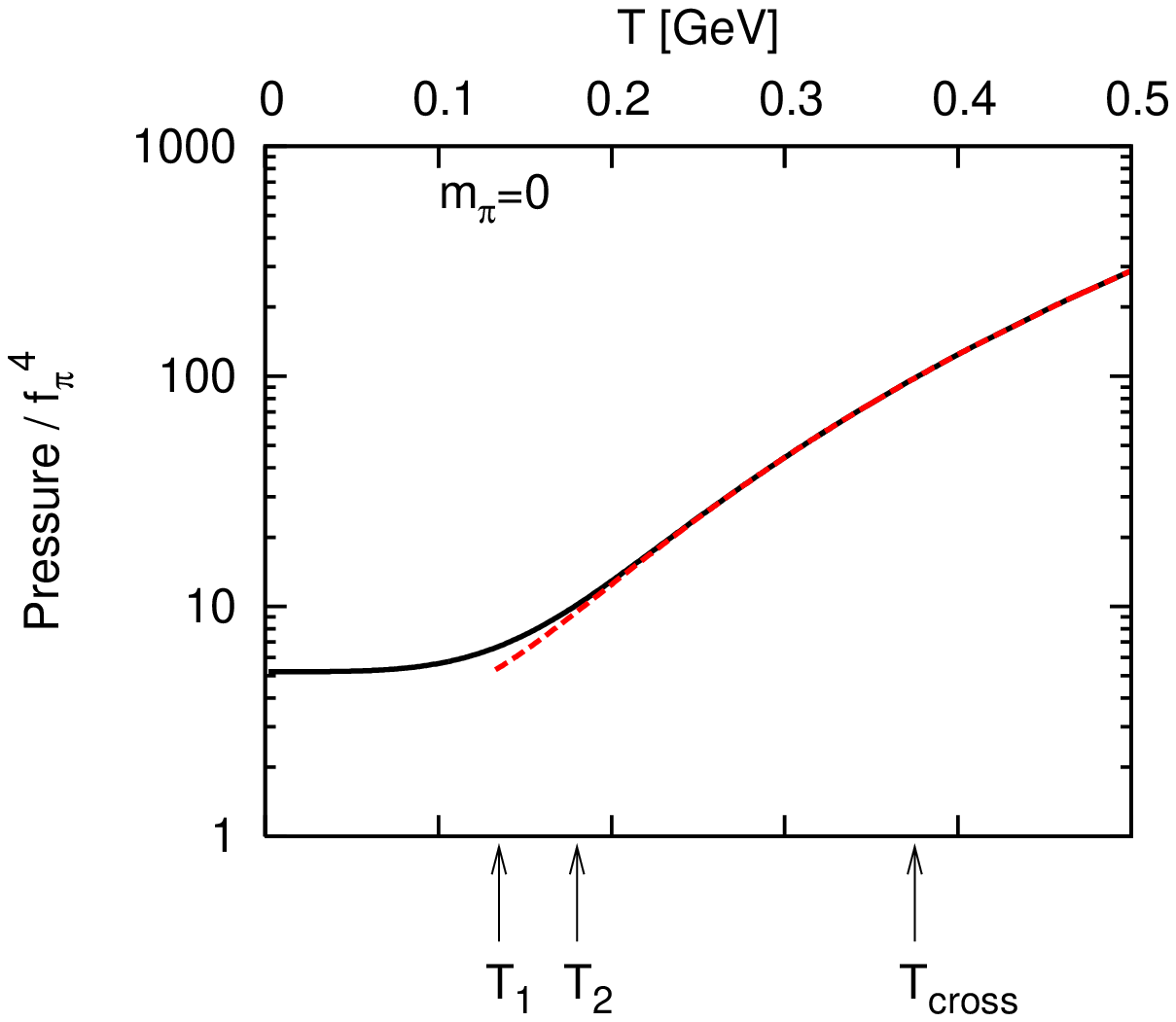}
\includegraphics[width=5.5cm,angle=-90]{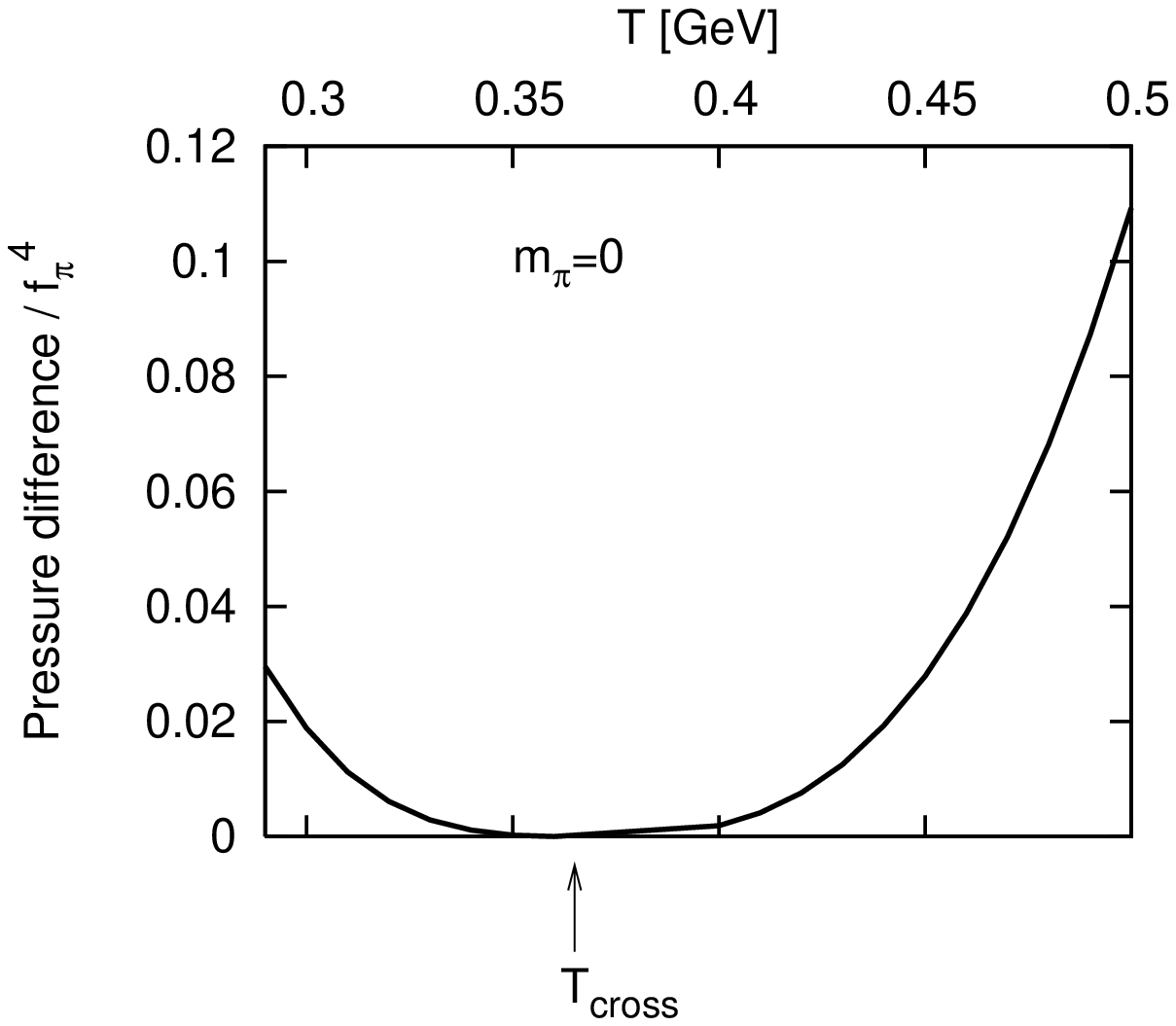}
\caption{Pressure (left panel) and pressure difference between 
  stable and metastable branches (right panel) as functions of
  temperature for $m_\pi=0$ case. The pressure for the stable and 
  metastable branches are presented by solid and dashed lines,
  respectively.}
\end{center}
\label{fig-v-mp0}
\end{figure*}

\subsection{Exact $O(4)$ Symmetry}

The results for stable and metastable branches in the case of exact
$O(4)$ symmetry, i.e. $H=0$, are displayed in Figs. 2 and 3. In fact,
it is impossible to resolve these branches near their crossing point
at $T_{\scr{cross}}\simeq$ 360 MeV.  However, these branches can be
distinctly resolved for the case of the partially violated
symmetry, see Figs. 4 and 5, which confirms our present identification.

The stable branch exists at all temperatures. It starts at $T=0$ from
the physical vacuum values of the masses and classical field and
crosses the metastable branch at $T_{\scr{cross}}$.  In terms of the
pressure, they are touching rather than crossing (see left panel of
Fig. 3).  Therefore, no transition from one branch to another occurs
at $T_{\scr{cross}}$. In the broken-symmetry phase, the stable branch behaves
precisely in accordance with our expectations -- the pion mass equals
zero.  Then a phase transition of the second order occurs at
$T_2\simeq$ 180 MeV, at which the field becomes zero. However, the
$\pi$ and $\sigma$ masses still differ beyond this transition point.
They become equal only after the second phase transition also of
second order at $T_{\scr{cross}}$. Note that the equal-mass solution
above $T_{\scr{cross}}$ is precisely the same as in the conventional
Hartree--Fock approximation, since the gapless modification term
(\ref{dPhi}) vanishes in this case.

The $T_1$ point proves to be irrelevant for the stable branch. Rather
it is the starting point for the metastable branch, which in the
range of $T_1<T<T_{\scr{cross}}$ precisely coincides with the solution
of the conventional Hartree--Fock approximation. The existence of this
branch above $T_{\scr{cross}}$ is an artifact of the introduced
NG-theorem restoring term $\Delta\Phi$. The corresponding field 
always vanishes for this branch.

Summarizing, in the spontaneously  broken-symmetry phase the NG
theorem is indeed fulfilled. Thus, we achieved the goal 
of the proposed modification of the Hartree--Fock approximation,
however partially, since the phase transition now proceeds through a
sequence of two second-order phase transitions rather than a single
one.

\begin{figure*}[ht]
\vspace*{5mm} 
\begin{center}
\includegraphics[width=5.5cm,angle=-90]{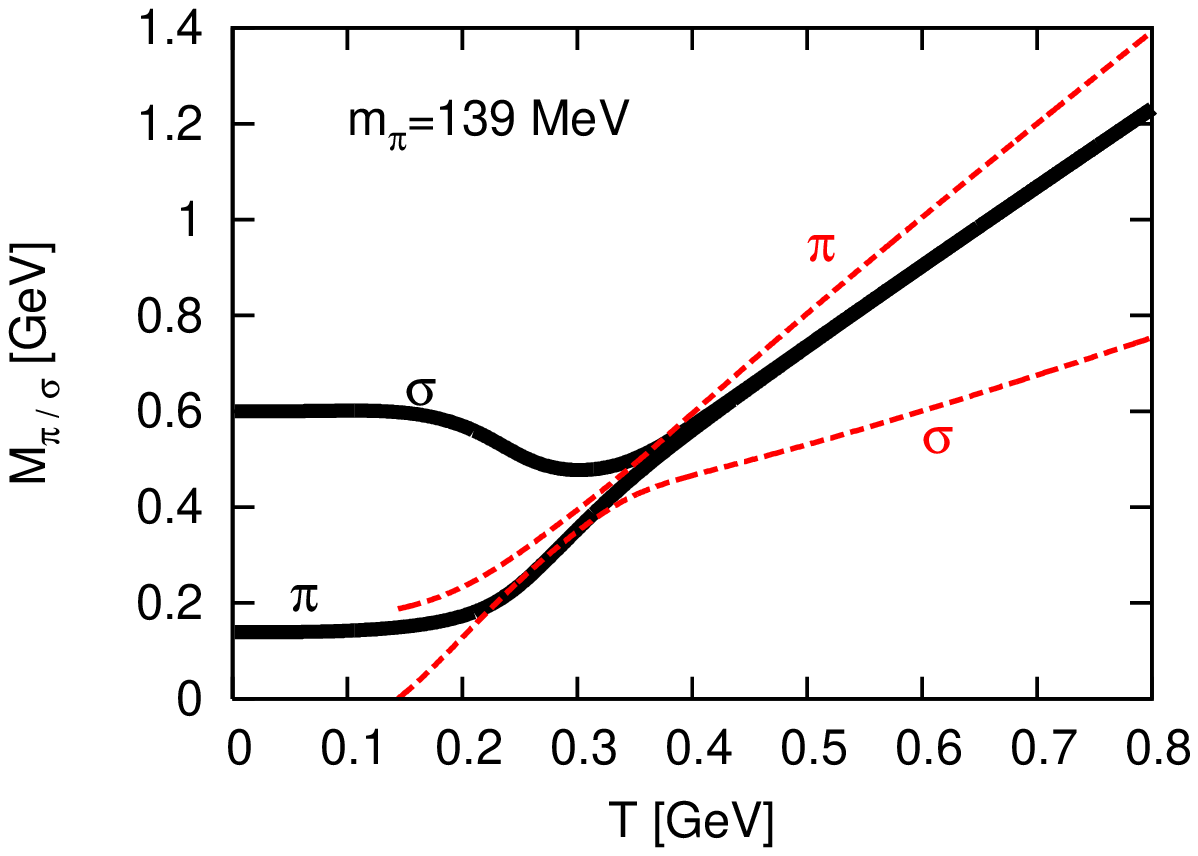}
\includegraphics[width=5.5cm,angle=-90]{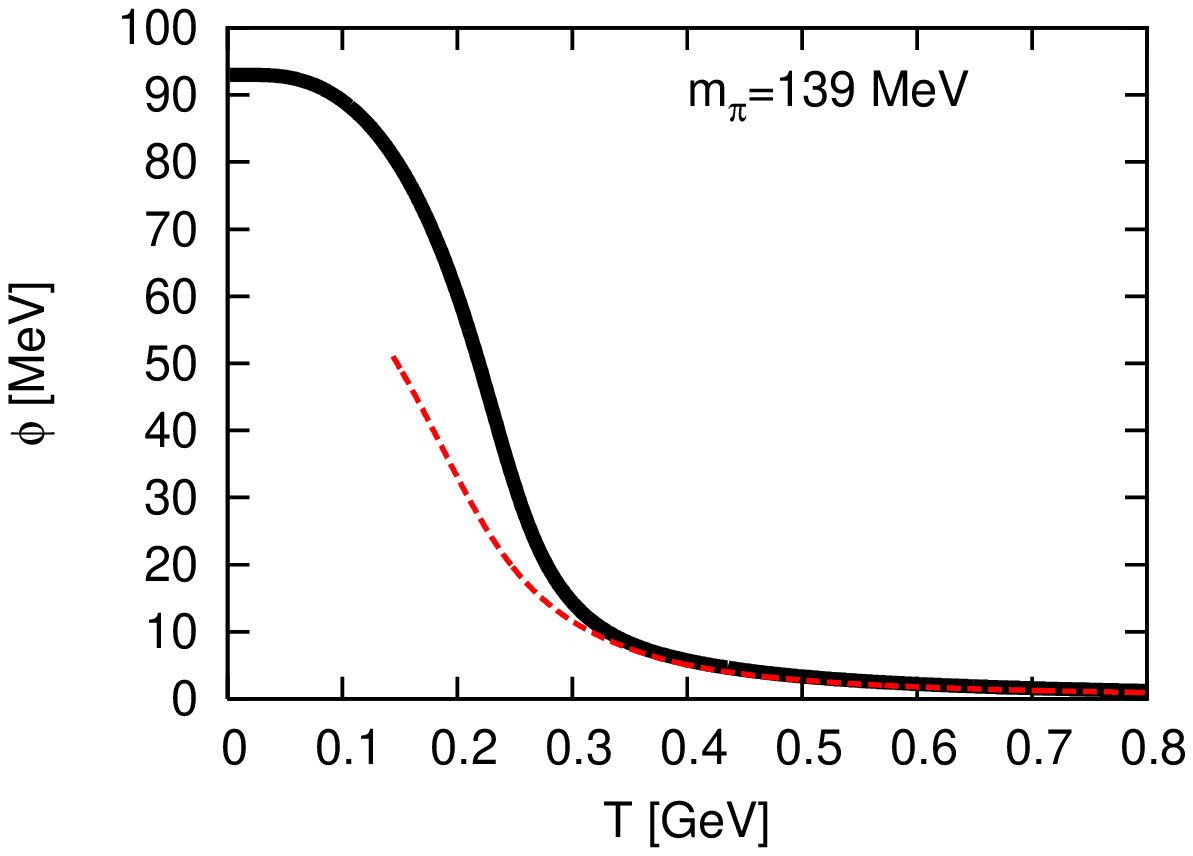}
\caption{Meson masses (left panel) and $\phi$ (right panel) as
  functions of temperature for $m_\pi=$ 139 MeV case (stable and
  metastable branches presented by solid and dashed lines, respectively).}
\end{center}
\label{fig-m-mp139}
\end{figure*}

\begin{figure*}[ht]
\begin{center}
\includegraphics[width=5.5cm,angle=-90]{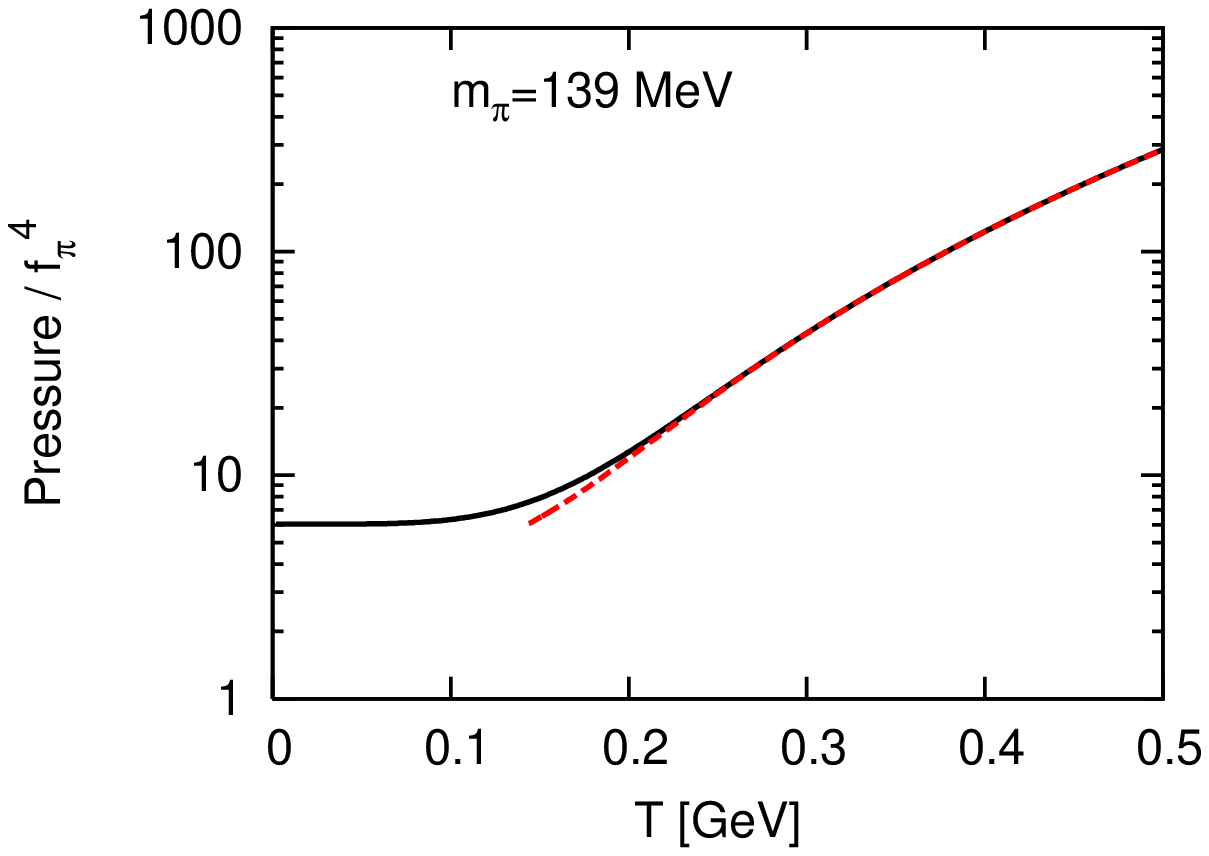}
\includegraphics[width=5.5cm,angle=-90]{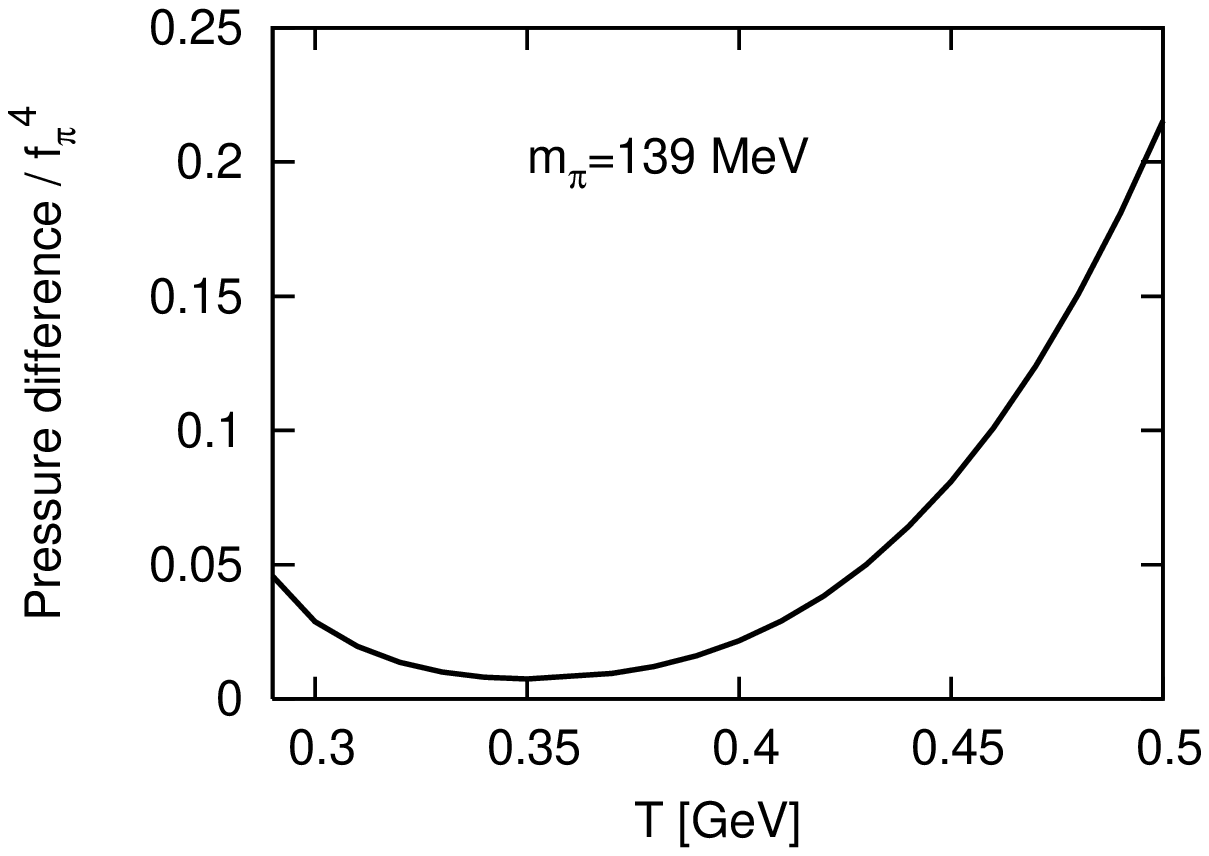}
\caption{Pressure (left panel) and pressure difference between 
  stable and metastable branches (right panel) as functions of
  temperature for $m_\pi=$ 139 MeV case (stable and metastable branches
  presented by solid and dashed lines, respectively). }
\end{center}
\label{fig-v-mp139}
\end{figure*}

\subsection{Approximate $O(4)$ Symmetry}

For the explicitly broken symmetry case (finite pion mass) the results for
both branches are displayed in Figs. 4 and 5.  In this case these two
branches are well resolved in the entire range of temperatures, since
they do not even touch each other, see Fig. 5 (right panel). As it was
expected, the sequence of two phase transitions is transformed here
into smooth cross-over transition.

\section{Conclusion}

A modified $\Phi$-derivable Hartree--Fock approximation to the
$O(N)$-model is proposed, which simultaneously preserves all the
desirable features of $\Phi$-derivable scheme (i.e. conservations and
thermodynamic consistency) while respecting the NG theorem in the
broken-symmetry phase. This is achieved by adding a correction $\Delta\Phi$ to
the conventional $\Phi$ functional. With this correction term, the
chiral phase transition proceeds through a sequence of two
second-order phase transitions rather than a single one.  In the first
transition the field disappears but the meson masses still remain
different. In the second transition also the masses become equal, and the
$O(N)$ symmetry is completely restored.

The nature of this correction can be understood as follows. For the
full theory, i.e. when all diagrams in $\Phi$ functional are taken
into account, the gapless and $\Phi$-derivable schemes are identical
and both respect the NG theorem. The conventional $\Phi$-derivable
Hartree--Fock approximation omits an infinite set of diagrams which
are necessary to restore its equivalence with the gapless scheme.
The $\Delta\Phi$ correction to the Hartree--Fock
approximation takes into account a part of those omitted diagrams
(at the level of actual approximation), and thus restores this equivalence
in the pion sector.

This $\Delta\Phi$ correction is unambiguously determined proceeding
from the following requirements: (i) it restores the NG theorem
in the broken-symmetry phase, (ii) it does not change
the Hartree--Fock equation for classical field, since the
$\Phi$-derivable and gapless schemes provide the same classical-field
equation already without any modifications, (iii) it does not change
results in the phase of restored $O(N)$ symmetry because there is no
need for it. 
If two last requirements are released, the  $\Delta\Phi$
correction become ambiguous: any 
\begin{widetext}
\begin{eqnarray}
\label{dPhi-alt}
&&\Delta\Phi_{\scr{alternative}} = 
-\frac{(N-1)\lambda}{2N} \left[(Q_\pi - Q_\sigma)^2
\right.
\cr
&+& 
\left.
f\left(\phi^2/(N-1)+Q_\pi-Q_\sigma,
  \phi^2/(N-1)+Q_\pi+Q_\sigma/(N-1)+const\right) \right], 
\end{eqnarray}
\newpage
\end{widetext}
where $f$ is an arbitrary function satisfying $\partial
f(x,y)/\partial x = 0$ at $x=0$, is as good as (\ref{dPhi}). The
arguments of the $f$ function are taken to be $O(N)$ invariants. 
For the first argument it was demonstrated in Eqs. (\ref{dPhi})
and (\ref{dPhi-ab}), whereas for the second argument it follows from
the identity $Q_{aa}=Q_\sigma+(N-1)Q_\pi$. Note that
our actual choice (\ref{dPhi}) is as close to the RPA result
\cite{HK3,Aouissat} as possible. 

The treatment of this paper was based on a naive renormalization
scheme, when all divergent parts of diagrams were simply omitted.
Presently, consistent renormalization schemes of $\Phi$-derivable
approximations are available \cite{HK3,HK1,HK2,Reinosa1,Reinosa2}. Our
first experience of dealing with them indicates that they give rise to
new (yet not always desirable) features of $\Phi$-derivable
approximations to theories with spontaneously broken symmetry. These
new features deserve special studies. To avoid mixing of effects of
the NG-theorem restoring term $\Delta\Phi$ with those of the
renormalization, we defer a discussion of the latter to a subsequent
paper.

The introduced gapless Hartree--Fock approximation
differs from approximations available at present. It differs from the
$1/N$ expansion, since it keeps all the terms in $N$, from the
Baym--Grinstein approximation \cite{Baym-Grin}, as it still is of the
$\Phi$-derivable nature, from gapless approximations
\cite{Aouissat,HK3}, etc. Of course, at the Hartree--Fock level there
is no ``collisional'' dissipation in the scheme (the Landau damping
may exist \cite{Cooper}). Nevertheless, the ``collisional''
dissipation can be introduced phenomenologically at the quasiparticle
level, i.e. by adding a collision term to the Vlasov kinetic equation
resulted from this modified Hartree--Fock approximation. Such an
approach looks preferable, e.g., for the DCC simulations
\cite{Randrup} as compared to that based on the Baym--Grinstein
approximation. The ``collisional'' dissipation naturally appears in
higher-order $\Phi$-derivable approximations, e.g. through the sunset
self-energy diagrams. The corresponding modifications of these
higher-order approximations in the way
described here are much more involved and will be reported elsewhere.

\acknowledgements

We are grateful to B. Friman, H. van Hees, D. Rischke, and
D.N. Voskresensky for 
useful discussions. One of the authors (Y.I.) acknowledges partial
support by the Deutsche Forschungsgemeinschaft (DFG project 436 RUS
113/558/0-2), the Russian Foundation for Basic Research (RFBR grant
03-02-04008) and
Russian Minpromnauki (grant NS-1885.2003.2).\\

\begin{widetext}

\appendix

{\large\bf Appendices}\\[-9mm]
\section{Tadpole}

In this appendix we present the limiting cases of the tad-pole
integral $Q^T$ introduced in Eq. (\ref{P}):
\begin{eqnarray}
\label{P2}
Q^T(M,T) = \frac{T^2}{2\pi^2} \int_{M/T}^\infty 
\frac{[\epsilon^2-(M/T)^2]^{1/2}}{e^\epsilon-1} d\epsilon
= \frac{1}{2\pi^2} M^2 \sum_{n=1}^\infty \frac{T}{nM} 
K_1\left(\frac{nM}{T}\right),  
\end{eqnarray}
where $\epsilon = (k^2+M^2)^{1/2}/T$. For $T\ll M$ 
\begin{eqnarray}
\label{Pinf}
Q^T(M,T \to 0) = (2\pi)^{-3/2} T^{3/2} M^{1/2} \exp(-M/T) ,  
\end{eqnarray}
while for  $M\ll T$
\begin{eqnarray}
\label{P20}
Q^T(M \to 0,T) = T^2 \left[
\frac{1}{12} - \frac{1}{4\pi} \frac{M}{T} - \frac{1}{8\pi^2} 
\left(\frac{M}{T}\right)^2 \ln\frac{M}{T}
\right]. 
\end{eqnarray}

\section{Second Derivative of CJT Effective Potential}

The equations of motion (\ref{Phi-SD-s})--(\ref{MF-eq}) in fact are
conditions which determine the extremum of the CJT effective
potential. However, stable solutions correspond only to minima of
this potential. To distinguish between minimum and maximum of the
potential, the sign of the second derivative of the potential over
$\phi$, under the condition that the Green functions solve the gap
equations with this $\phi$, should be determined. The first derivative
leads to
\begin{eqnarray}
\label{dVeff}
\frac{dV_{\scr{gHF}}(\phi,G(\phi))}{d\phi} = 
\frac{\partial V_{\scr{gHF}}(\phi,G)}{\partial \phi} + 
\frac{\delta V_{\scr{gHF}}(\phi,G)}{\delta G} 
\frac{dG}{d\phi} 
= \phi M^2_\pi - H,  
\end{eqnarray}
since
\begin{eqnarray}
\label{dV/dG}
\frac{\delta V_{\scr{gHF}}(\phi,G)}{\delta G} =0
\end{eqnarray}
due to equations of motion for Green functions. To obtain 
Eq. (\ref{dVeff}) we have used the equations of motion
(\ref{Phi-SD-p}) and (\ref{MF-eq}). Note that 
$dV_{\scr{gHF}}(\phi,G(\phi))/d\phi=0$, if $\phi$ is a solution to
Eq. (\ref{MF-eq}). Taking the next derivative one arrives at
\begin{eqnarray}
\label{d2Veff}
\frac{d^2V_{\scr{gHF}}(\phi,G(\phi))}{d\phi^2} 
= M^2_\pi +  \phi \frac{dM^2_\pi}{d\phi}.  
\end{eqnarray}
In order to find $dM^2_\pi/d\phi$ we can use the equations of motion
(\ref{Phi-SD-s}) and (\ref{Phi-SD-p}). Taking derivatives over $\phi$
in Eqs (\ref{Phi-SD-s}) and (\ref{Phi-SD-p}) and then resolving them
with respect to $dM^2_\pi/d\phi$, we arrive at 
\begin{eqnarray}
\label{dM2pi}
\frac{dM^2_\pi}{d\phi}=
\frac{\frac{\lambda}{N}2\phi M_\pi
\left[M_\sigma+\frac{\lambda}{N}2(N+2)R_\sigma\right]}%
{M_\sigma\left[M_\pi-\frac{\lambda}{N}(N-1)R_\pi\right]
+
\frac{\lambda}{N}R_\sigma
\left[(2N-5)M_\pi-\frac{\lambda}{N}2(N-1)(N+2)R_\pi\right]} 
\cr
\end{eqnarray}
with
\begin{eqnarray}
\label{R}
R_a = 
\frac{1}{2}\frac{\partial Q^T(M_a,T)}{\partial M_a}
\end{eqnarray}
To avoid uncertainties in the calculation of $\partial Q^T(M,T)/\partial M$
at $M\ll T$, it is reasonable to directly use Eq. (\ref{P20}). 
\end{widetext}


\end{fmffile}
\end{document}